\documentclass[letterpaper,prb,aps,floatfix,amsmath,onecolumn,nofootinbib,preprint]{revtex4}

\usepackage{amsmath}
\usepackage{bbm}
\usepackage{epsfig}
\usepackage{array}
\usepackage{verbatim}
\usepackage{hyperref}

\usepackage{amssymb}
\usepackage{multirow}
\usepackage{graphicx}

\usepackage{color}
\usepackage{ulem}

\newcommand{\beg}{\begin{equation}}

\newcommand{\en}{\end{equation}}

\newcommand{\eps}{\varepsilon}

\newcommand{\eref}[1]{Eq.~(\ref{#1})}

\renewcommand{\emph}{\textit}

\def\spvecA#1;{\if;#1;\else #1\cr \expandafter \spvecA \fi}

\graphicspath{ {images/} }
\usepackage{braket}

\begin{document}

\title{A note on Lee-Yang zeros in the negative half-plane\footnote{Dedicated to the memory of George Stell}}
\author{Joel L. Lebowitz$^{1,2}$, Jasen A. Scaramazza$^2$}
\date{\today}							

\affiliation{$^1$ Department of Mathematics, Rutgers University, Piscataway, NJ 08854, USA\\ $^2$ Department of Physics and Astronomy, Rutgers University, Piscataway, NJ 08854, USA}

\begin{abstract}
We obtain lower bounds on the inverse compressibility of systems whose Lee-Yang zeros of the grand-canonical partition function lie in the left half of the complex fugacity plane. This includes in particular systems whose zeros lie on the negative real axis such as the monomer-dimer system on a lattice. We also study the virial expansion of the pressure in powers of the density for such systems. We find no direct connection between the positivity of the virial coefficients and the negativity of the L-Y zeros, and provide examples of either one or both properties holding. An explicit calculation of the partition function of the monomer-dimer system on 2 rows shows that there are at most a finite number of negative virial coefficients in this case.
\end{abstract}

\maketitle

\clearpage

\section{Introduction}
The zeros of the grand canonical partition function (GPF) $\Xi(z,\Lambda)$, of equilibrium systems in a region $\Lambda$ at fugacity $z$, continue to be of interest [\onlinecite{L-YRev}] sixty years after their importance for identifying phase transitions was described by Lee and Yang [\onlinecite{L-Y1},\onlinecite{L-Y2}]. It turns out that in some simple models, the L-Y zeros are confined to the negative half $z$-plane, or even the negative real $z$-axis [\onlinecite{Hauge}-\onlinecite{Katsura}]. For example, Heilmann [\onlinecite{Heil}] showed that antiferromagnetic Ising models with pair interactions on line graphs (including, e.g., complete graphs) have L-Y zeros confined to the negative $z$-axis, which is a kind of antiferromagnetic analog to the circle theorem. For more recent work along this direction see [\onlinecite{LRS}] and references therein. 

Another recent study of an antiferromagnetic Ising-Heisenberg model on a diamond chain found that the nature of the distribution of L-Y zeros corresponds to distinct quantum ground states [\onlinecite{Anan}]. In the model considered, one ferrimagnetic phase corresponds to the L-Y zeros confined to the negative $z$-axis, while another ferrimagnetic phase corresponds to the L-Y zeros being both on the negative axis and on the unit circle. It was also shown recently that when the zeros of the GPF all lie in the left half of the complex $z$-plane, the system satisfies a local central limit theorem [\onlinecite{LF14},\onlinecite{LPRS}].

In this note we obtain some new results for the thermodynamic properties of systems with L-Y zeros confined to the negative half $z$-plane. We also discuss the relation between the location of the zeros and the sign of the virial coefficients of such systems. Along the way, we obtain an exact expression for the limiting behavior of the GPF for all systems with hard cores in the limit $z\to \infty$, $\Lambda$ finite.

\section{General properties of L-Y zeros}
\label{general}
For systems with superstable [\onlinecite{Ruelle}] Hamiltonians, which includes all systems with hard cores and interactions decaying fast enough, as well as ideal fermions, $\Xi(z,\Lambda)$, $\Lambda \subset \mathbb{R}^d$ (or $\Lambda \subset \mathbb{Z}^d$), can be written as a product over its roots, $z_{\alpha}\equiv-1/\eta_{\alpha}$ [\onlinecite{Ruelle}]
\beg
\Xi(z,\Lambda) = \prod_{\alpha=1}^{N_m}\left( 1 + \eta_{\alpha}z\right), \quad 1 \le N_m \le \infty,
\label{factorize}
\en
where $N_m=N_m(\Lambda)$ is the maximum number of particles that can be contained in $\Lambda$. We shall restrict ourselves here to systems with $N_m(\Lambda)<\infty$ and write $N_m=\rho_m |\Lambda|$, where $|\Lambda|$ is the volume of $\Lambda$ (or the number of lattice sites), without explicitly indicating the dependence of $\rho_m$ on $\Lambda$. The roots of $\Xi(z,\Lambda)$ are either negative or come in complex conjugate pairs and depend on $\Lambda$, the inverse temperature $\beta$, and the interactions - dependencies which we will not write out explicitly\footnote{The zeros will also depend for finite systems on the boundary conditions. This dependence disappears in the thermodynamic limit as can be seen by taking limits in \eref{PD2a} as is done in \eref{PD3}.}.

The pressure $p(z,\Lambda)$ is given by
\beg
\begin{split}
p(z,\Lambda)  &= \frac{\ln\left[ \Xi(z,\Lambda) \right]}{\left| \Lambda \right|} = \frac{1}{|\Lambda|}\sum_{\alpha=1}^{N_m}\ln\left( 1 + \eta_{\alpha}z\right)\\
&=\rho_m\biggl\langle  \ln(1+\eta z)\biggr\rangle.
\end{split}
\label{PD1}
\en
The angular brackets indicate an average over the $\eta_{\alpha}$: $\langle f(z,\eta)\rangle\equiv\frac{1}{N_m}\sum_{\alpha=1}^{N_m}f(z,\eta_{\alpha})$. Expanding \eref{PD1} in powers of $z$, we obtain the Mayer fugacity expansion [\onlinecite{PenroseFug}]
\beg
p(z,\Lambda)  = \sum_{j=1}^{\infty} b_j(\Lambda) z^j,
\label{PD2}
\en
where the $b_j(\Lambda)$ are given by
\beg
\begin{split}
b_j(\Lambda) &= \frac{(-1)^{j+1}}{j}\rho_m m_j(\Lambda),
\end{split}
\label{PD2a}
\en
with $m_j(\Lambda) \equiv \langle \eta^j \rangle$, the $j$-th moment of the $\eta_{\alpha}$'s.

Restricting our attention to regular [\onlinecite{Ruelle}] pair potentials $\phi(r)$, where $r$ is the particle separation, the $b_j(\Lambda)$ are given by the Mayer cluster integrals [\onlinecite{Stell}]. For example, the second cluster integral $b_2(\Lambda)$ is given by
\beg
\begin{split}
b_2(\Lambda) &= \frac{1}{2|\Lambda|}\int_{\Lambda}d\,\mathbf{r_2} \int_{\Lambda}(e^{-\beta \phi(r_{12})} - 1)d\,\mathbf{r_1},\\
\lim_{\Lambda\to\mathbb{R}^d}b_2(\Lambda) &= \frac{1}{2}\int_{\mathbb{R}^d}(e^{-\beta \phi(r)} - 1)d\,\mathbf{r}.
\end{split}
\label{PD3}
\en
For lattice systems, the integral is replaced by a sum.

The average density in the grand canonical ensemble is given by
\beg
\begin{split}
\rho(z,\Lambda) &= z \frac{d p(z,\Lambda)}{dz} =  \rho_m \biggl\langle \frac{\eta\,z}{1+\eta\,z} \biggr\rangle\\
&=  \sum_{j=1}^{\infty} j\,b_j(\Lambda) z^j.
\end{split}
\label{rhodefinition}
\en
The virial expansion is then obtained by eliminating $z$ between \eref{PD2} and \eref{rhodefinition} and writing
\beg
p(\rho,\Lambda)\equiv p(z(\rho,\Lambda),\Lambda)=\sum_{j=1}^{\infty} B_j(\Lambda)\rho^j.
\label{1virial}
\en
The relation between the $B_j(\Lambda)$ and $b_i(\Lambda)$, $i \le j$ was derived in [\onlinecite{Mayer}], but we will not make use of that here. We also do not consider the direct derivation of the virial expansion from the canonical partition function given in [\onlinecite{Pul}]. The latter differs from \eref{1virial} by terms which vanish in the thermodynamic limit $\Lambda \to \mathbb{R}^d(\mathbb{Z}^d)$. The rate of approach to equality between the different ensembles may depend on $z$ (see below).

The $|\eta_{\alpha}|$'s all lie in the range
\beg
\begin{split}
|z_m(\Lambda)|^{-1}\le |\eta_{\alpha}(\Lambda)|\le|z_0(\Lambda)|^{-1},
\end{split}
\label{rangeeta}
\en
where $z_0(\Lambda)$ ($z_m(\Lambda)$) are the zeros of $\Xi(z,\Lambda)$ with the smallest (largest) absolute value. Note that $|z_0(\Lambda)|=R(\Lambda)$, the radius of convergence of the fugacity series \eref{PD2}. There is a simple lower bound to $R(\Lambda)$ uniform in $|\Lambda|$ [\onlinecite{PenroseFug}, \onlinecite{Groeneveld},\onlinecite{RuelleB}] (for positive potentials it is $R(\Lambda)\ge \frac{1}{2e|b_2(\Lambda)|}$). Thus the $|\eta_{\alpha}(\Lambda)|$ remain bounded above when $|\Lambda|\to\infty$, but can approach zero in this limit. This fact and \eref{PD2a} ensures that there exists a limiting distribution for the $\eta_{\alpha}$'s in the thermodynamic limit, i.e. $\langle f(z) \rangle \to \int f(z,\eta) \nu(\eta)d\eta$.

The radius of convergence $R$ of the power series in $z$, obtained by interchanging the sum in \eref{PD2} and the limit $\Lambda \to \mathbb{R}^d$ satisfies $R \ge \lim_{\Lambda\to\mathbb{R}^d}\,R(\Lambda)$ [\onlinecite{PenroseFug}] with equality when $\phi(r) \ge 0$. There is also a lower bound for the radius of convergence $\mathcal{R}(\Lambda)$ of the virial expansion [\onlinecite{LebPen}], satisfying $\mathcal{R}(\Lambda)\le \mathcal{R}$, where $\mathcal{R}$ is the radius of convergence of the series \eref{1virial} when $B_j(\Lambda)$ is replaced with $B_j=\lim_{\Lambda\to\mathbb{R}^d}\,B_j(\Lambda)$.

\section{Results for L-Y zeros in the negative half plane}
\label{neghalfplane}
It follows from \eref{PD2a} that if the zeros all lie on the negative $z$-axis, the $b_j(\Lambda)$ alternate in sign. This alternation of signs also holds if $\phi(r)\ge0$ [\onlinecite{LiebAlt}] although $\phi(r)\ge0$ is not a necessary condition for the zeros to be on the negative $z$-axis [\onlinecite{Niemeyer}]. If $\phi(r)$ is negative over any finite range, however, then there exists a $\beta^*$ sufficiently large (i.e. a sufficiently low temperature) such that the alternation in sign does not hold. Therefore the L-Y zeros can only stay on the negative real z-axis for all temperatures if $\phi(r) \ge 0$.

Let $\eta_{\alpha}=x_{\alpha}+i\,y_{\alpha}$. Combining complex conjugate pairs in \eref{rhodefinition} leads to
\beg
\begin{split}
\rho(z,\Lambda) &=\rho_m \left( 1 - \biggl\langle n(z) \biggr\rangle\right),
\end{split}
\label{rhodefinition2}
\en
where the dependence of $n_{\alpha}(z)$ on $\eta_{\alpha}$ is given by
\beg
\begin{split}
n_{\alpha}(z) &\equiv \frac{1+x_{\alpha}z}{(1+x_{\alpha}z)^2+y_{\alpha}^2z^2}.
\end{split}
\label{rhodefinition3}
\en
The fluctuation in particle number is given by differentiating \eref{rhodefinition2}
\beg
\begin{split}
z\frac{d \rho(z,\Lambda)}{dz} &=\frac{1}{|\Lambda|}\left(\overline{N^2}(z,\Lambda)-\overline{N}^2(z,\Lambda)\right),\\
\end{split}
\label{Nvariance}
\en
where an overbar $\overline{\cdot}$ indicates the ensemble average over the grand canonical measure. We now write \eref{Nvariance} in terms of averages over $\eta_{\alpha}$
\beg
\begin{split}
z\frac{d \rho(z,\Lambda)}{dz} &= \rho(z,\Lambda)\left(1-\frac{\rho(z,\Lambda)}{\rho_m}\right) - \biggl(V(z,\Lambda)-W(z,\Lambda)\biggr),\\
\end{split}
\label{PD1222}
\en
where $V(z,\Lambda)$ and $W(z,\Lambda)$ are variances
\beg
\begin{split}
V(z,\Lambda) &= \rho_m\biggl\langle \left(n(z)-\bigl\langle n(z) \bigr\rangle\right)^2\biggr\rangle\ge0,\\
W(z,\Lambda) &= \rho_m\biggl\langle m^2(z) \biggr\rangle\ge0,\\
m_{\alpha}(z) &\equiv \frac{y_{\alpha}z}{(1+x_{\alpha}z)^2+y_{\alpha}^2z^2}.
\end{split}
\label{PD12222}
\en
where $\langle m(z) \rangle=0$ by symmetry of the L-Y zeros about the negative $z$-axis\footnote{Note that the quantity $V(z,\Lambda)-W(z,\Lambda)$ in \eref{PD1222} is a measure of the difference in variance of the L-Y zeros along the real and imaginary directions.}.

When the L-Y zeros are in the negative half plane, i.e. $x_{\alpha} \ge 0$,  it is helpful to rewrite \eref{PD1222} in the form
\beg
\begin{split}
z\frac{d \rho(z,\Lambda)}{dz} &= 2\rho(z,\Lambda)\left(1-\frac{\rho(z,\Lambda)}{\rho_m}\right) - 2V(z,\Lambda) - A(z,\Lambda),\\
A(z,\Lambda) &= \rho_m \biggl\langle \frac{x\, z}{(1+x\,z)^2+y^2z^2}\biggr\rangle \ge0\quad \textrm{if}\quad x_{\alpha}\ge0.
\end{split}
\label{PD122222}
\en
It follows from \eref{PD122222} that when the L-Y zeros are restricted to the negative half $z$-plane, there is a lower bound on the inverse compressibility
\beg
\begin{split}
\frac{dp(\rho,\Lambda)}{d\rho}\equiv \frac{\rho(z,\Lambda)}{z\frac{d\rho(z,\Lambda)}{dz}} \ge \frac{1}{2}\frac{1}{(1-\rho/\rho_m)}, \quad x_{\alpha}\ge0.
\end{split}
\label{PD1522}
\en
When the zeros all lie on the negative $z$-axis, $y_{\alpha}=0$, $x_{\alpha}=\eta_{\alpha}\ge0$, and $W(z,\Lambda)$ in \eref{PD1222} vanishes. We therefore obtain
\beg
\begin{split}
\frac{dp(\rho,\Lambda)}{d\rho} \ge \frac{1}{1-\rho/\rho_m}, \quad \eta_{\alpha}\ge0.
\end{split}
\label{PD15222}
\en
Furthermore, differentiating $\rho(z,\Lambda)$ with respect to $z$ when $\eta_{\alpha}\ge0$ we find
\beg
\frac{d^k\rho(z,\Lambda)}{dz^k}=(-1)^{k+1}\rho_m\biggl\langle \frac{\eta^k}{(1+\eta\,z)^{k+1}}\biggr\rangle,\quad k\ge1,\,\eta_{\alpha}\ge0,
\label{alterns}
\en
which alternates in sign with $k$.

The inequality \eref{PD15222} becomes an equality for the ideal lattice gas, when the only interaction is the hard core exclusion preventing the occupancy of any lattice site by more than one particle. In that case $\rho_m=1$, $\Xi(z,\Lambda)=(1+z)^{|\Lambda|}$, and all the L-Y zeroes are located at $z=-1$, i.e. $\eta_{\alpha}=1$ for all $\alpha$. Therefore $V(z,\Lambda)=0$ and $p(\rho,\Lambda)=-\ln(1-\rho)$. 

While Eqs.~(\ref{PD1522}-\ref{alterns}) remain valid in the thermodynamic limit, where $p(\rho)$ is the same for all ensembles, including the grand canonical and canonical ensembles, the same is not true of the limiting equality
\beg
\begin{split}
\lim_{z\to \infty}\left(1-\frac{\rho(z,\Lambda)}{\rho_m(\Lambda)}\right)\frac{dp(z,\Lambda)/dz}{d\rho(z,\Lambda)/dz}=1,\quad N_m(\Lambda)<\infty.
\end{split}
\label{limiteq}
\en
\eref{limiteq} follows from the fact that for finite $N_m(\Lambda)$, the $|\eta_{\alpha}|$'s are bounded away from zero, and is valid independent of the location of the L-Y zeros. Note, however, that \eref{limiteq} may fail if the thermodyamic limit is taken before $z\to\infty$ or when one uses the canonical ensemble definition of the pressure. Such is the case for the lattice systems with extended hard cores discussed in Sect.~\ref{latticegases}.

Another (more interesting) way of writing \eref{limiteq}, using \eref{Nvariance}, is
\beg
\begin{split}
\lim_{z\to \infty}\left\{ \frac{N_m(\Lambda)-\overline{N}(z,\Lambda)}{\overline{N^2}(z,\Lambda)-\overline{N}^2(z,\Lambda)} \right\}\frac{\overline{N}(z,\Lambda)}{N_m(\Lambda)}=1,\quad N_m(\Lambda)<\infty,
\end{split}
\label{limiteq2}
\en
where both the numerator and the denominator vanish as $z\to \infty$. \eref{limiteq2} also can be obtained by keeping only the terms proportional to $z^{N_m-1}$ and $z^{N_m}$ in the GPF when $z\to \infty$ at fixed $\Lambda$.

\section{The virial expansion}
\label{VE}
The thermodynamic properties of a gas are determined at small densities by the virial expansion (VE) of the pressure $p(\rho,\Lambda)$ in powers of the density $\rho$ given in \eref{1virial}. The low order terms in the expansion can be readily computed for classical systems with pair interactions $\phi(r)$ [\onlinecite{Stell}]. This can be done analytically or numerically in terms of the irreducible Mayer cluster integrals [\onlinecite{Pathria}]. In practice one only computes $B_j =\lim_{\Lambda \to \mathbb{R}^d(\mathbb{Z}^d)}B_j(\Lambda)$.

For the system of hard spheres (HS) in $\mathbb{R}^d$, which is the paradigm model for representing the effective strong repulsion between atoms at short distances, the $B_j$ are known in $d=3$ for $j\le 12$ [\onlinecite{wheats}], with high accuracy for $j \le 11$ (Boltzmann had computed the first four). In $d=2$, the $B_j$ are known for $j\le 10$ [\onlinecite{MC05}]. In $d=1$, $p=\frac{\rho}{1-\rho \,a}$, where $a=\rho_m^{-1}$ is the diameter of the hard rods, so that $B_j=a^{j-1}$, for all $j$. Remarkably, all known $B_j$ for $d=1,2,3$ are positive, which has led to the speculation that in fact all $B_j$ in $d=2,3$ are positive. It is known, however, that this is false in $d=5$, so that it is now generally expected that there will be negative $B_j$ in $d=3$, but perhaps not in $d=2$.

The physical interest in the positivity of the $B_j$ lies in the fact that one would like to extrapolate from the low density regime, well described by the first few terms in the virial expansion, to obtain information about $p(\rho) = \lim_{\Lambda \to \mathbb{R}^d}p(\rho,\Lambda)$ at higher densities, including possibly about the fluid-solid transition in $d=3$. Based on numerical simulations (Monte Carlo or Molecular Dynamics), one expects this transition to occur at a density of $\rho_f\approx .49/v$ where $v = \pi/6\, a^3$ is the actual volume occupied by a sphere with diameter $a$. The extrapolations of the pressure to higher densities take many forms and are in very good agreement with the machine results for $\rho \lesssim \rho_f$ [\onlinecite{Mulero}-\onlinecite{Goos}]; in particular, see Ref. [\onlinecite{Goos}] and references there for highly accurate results. Some even give very high accuracy results for the metastable extension between $\rho_f$ and the random close-packing density $\rho_r \approx .64/v$, a region which may also contain a transition from metastable fluid to glass [\onlinecite{Jadrich,Parisi,GlassPhys}].

In many of these (approximate) extrapolations, the radius of convergence $\mathcal{R}$ of the virial expansion is determined by a singularity at some positive value of the density $\tilde{\rho}=\mathcal{R}$, with $\tilde{\rho}>\rho_f$ in $d=3$ [\onlinecite{Robles}-\onlinecite{Goos}]. This will certainly hold when all $B_j$, or all but a finite number of them, are positive, but need not be the case otherwise. In fact, for hard hexagons on a triangular lattice $\mathcal{R}$ is determined by a singularity of $p(\rho)$ at $\rho '$, with $\rho '$ complex [\onlinecite{Gaunt}] and smaller in modulus than the disorder-order transition $\rho_d$ [\onlinecite{Baxter}], $|\rho '|<\rho_d$. If this were true also for hard spheres in $d\ge2$ it would limit the utility of extrapolating the virial expansion beyond the rarified-gas phase.

Here we consider the relation between the signs of the $B_j$ and the location of the Lee-Yang zeros in the complex $z$-plane. All the previously known examples of almost all positive (i.e. a finite number of negative) $B_j$ were for systems for which all the L-Y zeros lie on the negative $z$-axis. This behavior fits in with the conjecture by Federbush, et al., that all the $B_j$ for the monomer-dimer system on regular lattices are positive [\onlinecite{Federbush}], since there the L-Y zeros are indeed on the negative $z$-axis [\onlinecite{Lieb}]. Systems with strictly negative L-Y zeros do not have any phase transition, but this does not rule out the possibility that a system with a phase transition has almost all $B_j\ge0$.

In fact, we do not find a direct connection between the negativity of the L-Y zeros and the possibility of almost all positive virial coefficients, and indeed we find models that have only the former property, only the latter, or both. This negative result leaves open the positivity of almost all positive virial coefficients for hard spheres in $d=2$ or $d=3$ ($\mathcal{R}$ could even be larger than the close packing density). There does seem, however, to be some connection between the proximity of L-Y zeros to the negative axis and the positivity of virial coefficients (see examples below) for which we have no complete explanation at the present time. For example, in terms of the moments $m_j=\langle\eta^j\rangle$ introduced in \eref{PD2a}, the first few $B_j$ are
\beg
\begin{split}
B_1 &= 1 = \rho_m\,m_1,\\
B_2 &= \frac{\rho_m}{2}m_2,\\
B_3 &= \rho_m^2m_2^2-\frac{2}{3}\rho_m\,m_3.
\end{split}
\label{firstfew}
\en
It follows that $B_2>0$ iff $\langle x^2 \rangle > \langle y^2\rangle$, where the $x_{\alpha}$ and $y_{\alpha}$ are defined in Sect.~\ref{neghalfplane} as the real and imaginary parts of the negative inverse L-Y zeros $\eta_{\alpha}$, respectively.

\subsection{Hard core lattice gases in 1D}
\label{latticegases}

(i) Consider a 1D lattice of $N$ sites separated by unit distance (the lattice length is $L = N$) with the pair potential $u(x_{ij})$, ($x_{ij}=|i-j|$)
\beg
\begin{split}
u(x_{ij}) &= 0 \textrm{ if $x_{ij} \ge q$}, \\
u(x_{ij}) &= \infty \textrm{ if $x_{ij} < q$}.
\end{split}
\label{LG}
\en
The integer $q$, $q\ge1$, is called the ``exclusion factor" ($q=1$ is the ideal lattice gas and $q=2$ is isomorphic to the 1D monomer-dimer problem). Using the canonical ensemble, it was shown in Ref.~[\onlinecite{L-Y2}] that for $L\to\infty$
\beg
\begin{split}
p(\rho) &= \ln \left(1 +  \frac{\rho}{1 - \rho/\rho_m }\right)\\
&= -\ln(1-\rho/\rho_m)+\ln(1-\rho/\rho_m+\rho),
\label{pLG}
\end{split}
\en
where $\rho_m=1/q$. Note that $\lim_{\rho\to\rho_m}(1-\rho/\rho_m)\frac{dp(\rho)}{d\rho}=1/\rho_m\ne1$ for $q>1$, which gives an example of \eref{limiteq} failing when the thermodynamic limit is taken before letting $z\to\infty$. Expanding \eref{pLG} we see that
\beg
B_j = \frac{1-(1-\rho_m)^j}{\rho_m^j\,j}>0,\,\forall j,
\en
which agrees with the finite-$L$ grand canonical virial coefficients from \eref{1virial} only in the thermodynamic limit.
If we introduce the lattice constant $\delta$ and take the continuum limit $q\,\delta\to a$ as $q\to \infty$, the problem reduces to that of hard rods of length $a$ on a line with pressure
\beg
\label{pLG2}
\begin{split}
p(\rho) &= \frac{\rho}{1 - \rho/\rho_m}.
\end{split}
\en
where $\rho_m=1/a$. The L-Y zeros of the lattice model are on the negative real axis [\onlinecite{Hauge2}], as are those of the continuum model [\onlinecite{Elvey}]. Note however the change in the nature of the singularity as $\rho\to\rho_m$, with the divergence in the continuum case much stronger than in the lattice.

One can also derive the distribution of the L-Y zeros when $L\to\infty$ for the $q$-exclusion models. The equation of state, \eref{pLG}, gives $z(p)$
\beg
\begin{split}
z(p) = (e^p-1)e^{p(q-1)},
\label{zLG}
\end{split}
\en
Upon using the Lagrange inversion formula to obtain the moments defined in \eref{PD2a}, $m_j=\langle \eta^j\rangle$, we find
\beg
\begin{split}
m^{[q]}_j = \lim_{\Lambda \to \infty}m^{[q]}_j(\Lambda) = \binom{j\,q}{j}, \quad j\ge 0.
\label{zLG2}
\end{split}
\en
For $q=1$, the $m^{[1]}_j=1$ are moments of a delta function $\delta(\eta-1)$, which corresponds to the ideal lattice gas. The authors of Ref.~[\onlinecite{Polska}] consider a more general set of binomial sequences $m^{[t,r]}_j=\binom{tj+r}{j}$. They find that the $m_j^{[t,r]}$ are moments of a probability density function $h_{t,r}(\eta)$ with support on a domain $\mathcal{D}_t\subseteq [0,t^t(t-1)^{1-t}]$. Here we have $t=q\in \mathbb{N}$, $r=0$, and $h_{q,0}(\eta)$ can be written in terms of the Meijer G-function [\onlinecite{Polska}]. This permits, for some $q$, to express the limiting density of the $\eta_{\alpha}$'s in terms of elementary functions. For example, for $q=2$
\beg
\begin{split}
h_{2,0}(\eta) &= \frac{1}{\pi\sqrt{\eta(4-\eta)}},\quad \eta\in(0,4),
\label{zLG22}
\end{split}
\en
as also derived in [\onlinecite{Hauge2}]. The density $h_{3,0}(\eta)$ also has a simple expression in terms of elementary functions for $\eta\in(0,27/4)$, which also diverges at the endpoints. The support of these divergences imply that the L-Y zeros of these two models reach infinity in the thermodynamic limit and that there is an accumulation of zeros near the smallest magnitude L-Y zero $z_0$.

(ii) The next model we consider is a lattice consisting of two rows in which particles exclude their nearest neighbor sites: two horizontal and one vertical. Using open boundary conditions $\Xi_{2N}(z)$ can be obtained using a transfer matrix $M$
\beg
\begin{split}
\Xi_{2N}(z)&=\mathbf{u}\cdot M^{N-1}\cdot \mathbf{v}^T_2,\\
\mathbf{u}&= \left(1,1,1\right),\quad \mathbf{v}_2=\left(1,z,z\right),\\
M &= \left(
\renewcommand{\arraystretch}{.25}
 \begin{array}{ccc}
1 & 1 & 1\\
 z & 0 & z\\
z & z & 0
\end{array} \right).
\end{split}
\label{asdf}
\en
The fugacity $z$, as a function of pressure $p$ in the $N\to\infty$ limit, is given by
\beg
\begin{split}
z(p)&= e^{2p} \tanh{p}.
\end{split}
\label{diamond}
\en
To locate the L-Y zeros of this model in the thermodynamic limit, we first diagonalize the matrix $M$ and expand $\Xi_{2N}(z)$ in terms of the eigenvalues. A necessary condition for $\Xi_{2N}(z)$ to vanish as $N\to \infty$ is $|\lambda_+(z)|=|\lambda_-(z)|$, where $\lambda_{\pm}(z)=\frac{1}{2} \left(\pm\sqrt{z^2+6 z+1}+z+1\right)$. The only $z$ that satisfy this relation are real and negative. Therefore, as $N\to\infty$, the L-Y zeros lie on the negative $z$-axis.

To locate the singularities in the $\rho$ plane, we get $p(\rho)$ from \eref{diamond}
\beg
\begin{split}
p(\rho)=-\rho_m\ln\left(1-\frac{\rho}{\rho_m}\right)+\rho_m\ln\left(\frac{\rho}{\rho_m}+\sqrt{1-2\frac{\rho}{\rho_m}+2\left(\frac{\rho}{\rho_m}\right)^2}\right).
\end{split}
\label{diamond2}
\en
where $\rho_m=1/2$. The nearest singularities of $p(\rho)$ occur at the square-root singularity in \eref{diamond2}, i.e. at $\rho_{\pm}=\frac{1}{4}(1\pm i)$. Thus there are an infinite number of negative virial coefficients. Though we limited our analysis to the 2-row case, [\onlinecite{Nil}] studies the location of the L-Y zeros of a nearest neighbor exclusion model
wound on a cylinder of circumference $k$ and infinite height. It is found that for $k>2$, the L-Y zeros move off the negative real axis.

(iii) The 2-row monomer-dimer system is considered in Appendix~\ref{appe}. We prove that it has almost all $B_j\ge0$. This is not as strong as what Federbush, et al conjecture, but it goes in that direction [\onlinecite{Federbush}].

\subsection{Square well interactions}
In all examples so far, systems with positive virial coefficients also have all L-Y zeros on the negative $z$-axis. Next we give an example where almost all (possibly all) virial coefficients are positive, but the L-Y zeros are off axis. Consider the 1D lattice gas with a nearest neighbor pair potential $\eps$. This system is isomorphic to the 1D Ising model with nearest neighbor interactions. The equation of state in the thermodynamic limit is (see [\onlinecite{Katsura}], for example)
\beg
\begin{split}
p(\rho) &= -\ln[1-\rho]+\ln\left[1- \frac{1-2 \alpha  \rho -\sqrt{1-4 \alpha  (1-\rho ) \rho }}{2 (1-\alpha )}\right] \\
\alpha &= 1-e^{-\eps}.
\label{EoS2inf}
\end{split}
\en
For $\alpha \ge 0$, the system is ferromagnetic and the L-Y zeros lie on a circle in the complex plane by the well-known L-Y circle theorem [\onlinecite{L-Y2}]. When $\alpha < 0$ the system is antiferromagnetic and the L-Y zeros are on the negative real axis with a known distribution [\onlinecite{Katsura}].

The virial coefficients $B_j$, however, are almost all positive over a range of both positive and negative potentials. The power series of the first logarithm in \eref{EoS2inf} has all positive coefficients and a radius of convergence of $\mathcal{R}_1=\rho_m=1$. If the series expansion of the second logarithm in \eref{EoS2inf} has a radius of convergence $\mathcal{R}_2>\mathcal{R}_1$, then it follows that its coefficients will eventually be smaller in magnitude than those of the first logarithm, and only a finite number of virial coefficients can be negative.

All singularities in the second logarithm occur when the square root in \eref{EoS2inf} vanishes. One finds
\beg
\begin{split}
\mathcal{R}_2 &= \mathcal{R}_2(\alpha), \\
&= \inf_{\pm} \left|\frac{1}{2}\left(1\pm\sqrt{1-\frac{1}{\alpha }}\right)\right| .
\label{R2}
\end{split}
\en
Note that for all $\alpha < 1$, $\mathcal{R}_2$ occurs at a complex value of $\rho$. For $|\alpha| \ll 1$, $\mathcal{R}_2>\mathcal{R}_1$ and the virial coefficients are almost all positive. As we increase the magnitude of $\alpha$, $\mathcal{R}_2$ decreases until it falls below $\mathcal{R}_1$, at which point there are an infinite number of negative virial coefficients. Only when $\alpha = 1$, the hard core limit, does $\mathcal{R}_2$ occur at a positive value of $\rho$. Defining $\alpha^*$ by $\mathcal{R}_2(\alpha^*) = \mathcal{R}_1$, we find
\beg
\begin{split}
\alpha^* &= -\frac{1}{8}, \quad \alpha < 0, \\
\alpha^* &= \frac{1}{4}, \quad \alpha > 0. \\
\label{R22}
\end{split}
\en
\eref{R22} shows that there is a range of positive and negative interactions over which all but a finite number of virial coefficients are positive, i.e.
\beg
-\ln{\frac{9}{8}} < \eps < \ln{\frac{4}{3}}.
\label{range}
\en
With numerical expansions of the equation of state \eref{EoS2inf}, we find that within the range of \eref{range} (including the endpoints), there are in fact no negative virial coefficients up to $\mathcal{O}(\rho^{2000})$.

Acknowledgements: We thank Eric Carlen, David Huse and Eduardo Waisman for helpful comments. We are particularly thankful to Gene Speer for great help with all aspects of this work. This work was supported by NSF grant DMR 1104500 and AFOSR Grant \#FA9550-16-1-0037.
\appendix
\section{Virial coefficients for the 2-row monomer-dimer problem}
\label{appe}
Heilmann and Lieb [\onlinecite{Lieb}] proved that monomer-dimer (MD) systems have all L-Y zeros on the negative real axis of the dimer fugacity $z$. It is conjectured in [\onlinecite{Federbush}] that the virial coefficients are always positive for MD systems on regular lattices. We prove here that this is indeed the case for the 2-row infinite square lattice (more precisely, we prove that only a finite number of coefficients can be negative).

We use a standard transfer matrix formalism to obtain the grand partition function on a $2\times N$ lattice with open boundary conditions. Any configuration of the system has one of five right-end states, corresponding to either no dimers, a single vertical dimer, a single horizontal dimer in the first or second row, or two horizontal dimers occupying both lattice sites on the far right-end of the lattice.
Let $\mathbf{v}_{N}$ be the vector whose $j$-th component is the GPF for configurations on a $2\times N$ lattice restricted to having the $j$-th right-end state. The total GPF $\Xi_{N}(z)$ is then
\beg
\label{MD2rowGPF1}
\begin{split}
\Xi_{N}(z) &= \mathbf{u}\cdot \mathbf{v}_{N}^T\\
&= \mathbf{u} \cdot M^{N-1}\cdot \mathbf{v}_{1}^T,
\end{split}
\en
where
\beg
\begin{split}
\mathbf{v}_{1} &= \left(1,z,0,0,0\right),\\
\mathbf{u} &= \left(1,1,1,1,1\right),
\end{split}
\en
and the transfer matrix $M$ is
\beg
\label{TransferMatrixExpl1}
\begin{split}
M &= \left(
\renewcommand{\arraystretch}{.25}
 \begin{array}{ccccc}
1 & 1 & 1 & 1 & 1\\
 z & z & z & z & z\\
z & 0 & 0 & z & 0 \\
z & 0 & z & 0 & 0 \\
z^2 & 0 & 0 & 0 & 0
\end{array} \right).
\end{split}
\en
For $z>0$, one can verify that $M^2$ has strictly positive entries, which implies by the Perron-Frobenious theorem that there is a unique, real, largest modulus eigenvalue of $M$ for $z>0$. Let $\lambda_m(z)$ be the largest eigenvalue of $M$ for $z\ge0$, then
\beg
\label{MDpd}
\begin{split}
p(z) &= \lim_{N\to \infty}\frac{1}{2N}\ln \Xi_{N}(z), \\
&= \frac{1}{2}\ln \lambda_m(z),\quad z\ge0.\\
\end{split}
\en
The characteristic polynomial $P(z,\lambda)$ of $M$ satisfies the equation
\beg
\label{CharPoly}
\begin{split}
P(z,\lambda) = -\lambda(\lambda+z)(\lambda^3-(1+2z)\lambda^2-z\lambda+z^3)=0,
\end{split}
\en
which implies that $\lambda_m(z)$ is the largest solution of
\beg
\label{CharPoly2}
\begin{split}
\lambda^3-(1+2z)\lambda^2-z\lambda+z^3=0.
\end{split}
\en

To analyze the series $p(\rho) = \sum_{j=1}^{\infty}B_j \rho^j$, where $\rho$ is the dimer density, we make a helpful change of variables to $t(z) \equiv \frac{z}{\lambda_m(z)}$. Using \eref{CharPoly2}, we find
\beg
\label{MDpd2}
\begin{split}
\textrm{for }&t \in [0,\phi^{-1}):\\
z(t)&=\frac{t(1+t)}{(1-t)(\phi^{-1}-t)(\phi+t)},\\
p(t) &= \rho_m\ln \left(\frac{1+t}{(1-t)(\phi^{-1}-t)(\phi+t)}\right),\\
\rho(t) &= \rho_m\left(1 - \frac{(1-t^2)(\phi^{-1}-t)(\phi+t)}{P_4(t)}\right),\\
P_4(t) &\equiv 1+2t-2t^2-2t^3-t^4.
\end{split}
\en
where $\phi$ is the golden mean, $\phi = (1+\sqrt{5})/2\approx 1.6108$, and $\rho_m=1/2$ is the maximum density.

\begin{figure}
\includegraphics[]{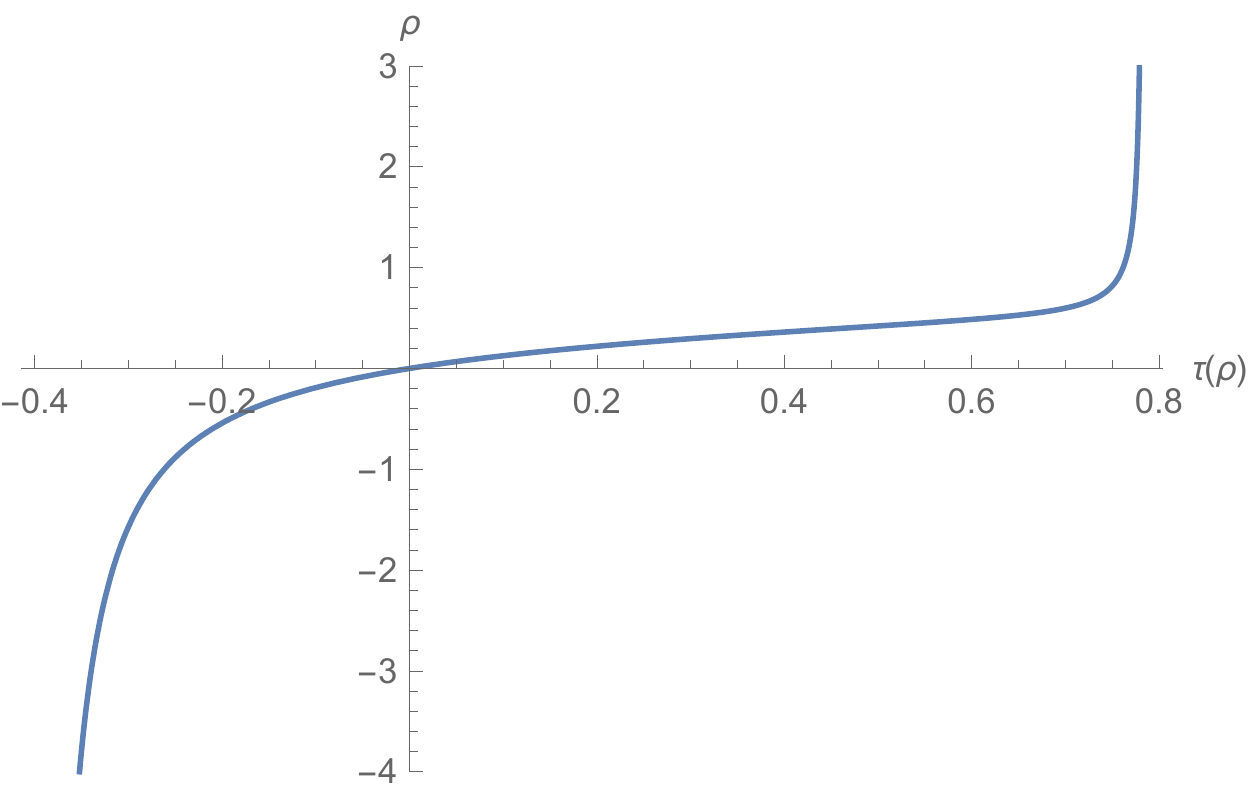}
\caption{The function $\tau(\rho)$ plotted on the horizontal axis. Note that $\tau(\rho)$ is a bijection from $\mathbb{R}$ to $\mathbb{T} \approx (-.394,.784)$.}
\label{Tbranch}
\end{figure}

Let $t_{1,2}$ be the two real roots of the polynomial $P_4(t)$, where $t_1\approx -.394$ and $t_2\approx .784$. On $\mathbb{T}=(t_1,t_2)$, $\rho(t)$ is a strictly increasing function, with $\rho(t)\to-\infty$ as $t\to t_1$ and $\rho(t)\to\infty$ as $t\to t_2$, so that we may define $\tau(\rho)$ to be the particular inverse function of $\rho(t)$ which takes values in $\mathbb{T}$. $\tau(\rho)$ is then a bijection from $\mathbb{R}$ to $\mathbb{T}$ (see Fig.~\ref{Tbranch}), with $\tau(\rho(t))=t$ for $t\in\mathbb{T}$. 
This definition of $\tau(\rho)$ allows us to define $p(\rho)$ on $\rho \in (-\infty,\rho_m)$
\beg
\label{MDpr}
\begin{split}
p(\rho) &= -\rho_m\ln\left(1-\rho/\rho_m\right)+\rho_m\ln\left(\frac{(1+\tau(\rho))^2}{P_4(\tau(\rho))}\right),\\
&= -\rho_m\ln\left(1-\rho/\rho_m\right)+f(\rho),\quad \rho \in \left(-\infty,\rho_m \right).
\end{split}
\en
We now analytically continue $p(\rho)$ from the domain $\rho \in (-\infty,\rho_m)$ into the complex plane in order to determine the location of its finite-$\rho$ singularities. In particular, the first logarithm in \eref{MDpr} is singular only at $\rho_m$ or $|\rho|\to \infty$. This term has all positive expansion coefficients about $\rho=0$ and a radius of convergence of $\mathcal{R}_1=\rho_m$. We will show that $f(\rho)$ is analytic inside a disk of radius $\mathcal{R}_2>\mathcal{R}_1$. This implies that asymptotically the coefficients in the expansion of $f(\rho)$ are smaller in magnitude than those coming from the first term, and therefore there are at most a finite number of negative virial coefficients.

The singularities of the secofnd logarithm in \eref{MDpr} occur either where $\tau(\rho)$ is singular or where $f(\rho)$ diverges. To discuss both of these possibilities it is convenient to define a single-valued ``physical'' branch of $\tau(\rho)$ in an appropriately cut complex $\rho$ plane.  We begin with $\tau(\rho)$ defined on the entire real axis (see Fig.~\ref{Tbranch}) and real analytic there. All possible singularities (branch points) $t^*$ of $\tau(\rho)$ may be found by solving $\frac{d\rho(t)}{dt}\bigl|_{t^*}=0$; using Mathematica to get numerical values, we find six solutions of this equation, with corresponding branch points in the $\rho$ plane given by $0.313\pm0.536\,i$, $0.497\pm0.121\,i$, $0.438$, and $1.039$. We now introduce four cuts in the $\rho$ plane, one beginning at each of the complex branch points and extending radially to $\infty$. The physical branch of $\tau(\rho)$ is defined in this plane by analytic continuation from the real axis, and is single valued.  Note that we do not need to consider the two real branch points since this physical branch is known to be analytic on the real axis; these branch points lie on other sheets of $\tau(\rho)$.

The singularities we must consider in determining ${\cal R}_2$, the radius of convergence of the power series for $f(\rho)$ in powers of $\rho$, are thus the four complex branch points defined above and the points where $f(\rho)$ diverges.  For the latter, we note that $f(\rho)$ diverges if $\tau(\rho)=-1$.  Using \eref{MDpd2}, however, we see that $\tau(\rho)=-1$ implies that $\rho=\rho_m$, and we know that on the physical branch, $\tau(\rho_m)=\phi^{-1}\ne-1$; thus this singularity cannot occur on the physical branch.  The only other way that $f(\rho)$ diverges is if $P_4(\tau(\rho))=0$.  Here we again use \eref{MDpd2} and we see that $|\rho(t)|=\infty$ if $P_4(t)=0$; these singularities occur at  $\infty$ in the $\rho$ plane and cannot affect the value of ${\cal R}_2$.

Since each of the complex branch points of $\tau(\rho)$ has absolute value greater than $\rho_m$, we conclude that ${\cal R}_2>{\cal R}_1$.  It follows that for $j\to\infty$, the $B_j$ are dominated by those obtained from the first logarithm in \eref{MDpr}. Hence, there are at most a finite number of negative virial coefficients in the virial expansion.

The solution of the finite \textit{periodic} 2-row lattice (not detailed here) yields the first $N-1$ infinite lattice virial coefficients exactly (see [\onlinecite{small}] for example). Numerically we checked the first 200 virial coefficients are positive, and we believe that the proven finite number of negative coefficients is indeed zero.

\end{document}